\documentclass[12pt]{iopart}
\usepackage{iopams}
\usepackage{graphicx}
\usepackage{cite}

\begin{document}
\title{Exceptional points and unitary evolution of the physical solutions}
\author{E. Hern\'andez$^{1}$, A. J\'auregui$^{2}$, D. Lohr$^{1}$ and
  A.  Mondrag\'on$^{1}$ } \address{$^{1}$Instituto de F\'{\i}sica,
  Universidad Nacional Aut\'onoma de M\'exico, Apdo. Postal 20-364,
  01000 M\'exico D.F. M\'exico} \address{$^{2}$Departamento de
  F\'{\i}sica, Universidad de Sonora, Apdo. Postal 1626, Hermosillo,
  Sonora, M\'exico} \ead{queta@fisica.unam.mx}

\begin{abstract}
An example of exceptional points in the continuous spectrum of a real,
pseudo-Hermitian Hamiltonian of von Neumann-Wigner type is presented
and discussed. Remarkably, these exceptional points are associated
with a double pole in the normalization factor of the Jost
eigenfunctions normalized to unit flux at infinity. At the exceptional
points, the two unnormalized Jost eigenfunctions are no longer
linearly independent but coalesce to give rise to two Jordan cycles of
generalized bound state eigenfunctions embedded in the continuum and a
Jordan block representation of the Hamiltonian. The regular scattering
eigenfunction vanishes at the exceptional point and the irregular
scattering eigenfunction has a double pole at that point. In
consequence, the time evolution of the regular scattering
eigenfunction is unitary, while the time evolution of the irregular
scattering eigenfunction is pseudounitary. The scattering matrix is a
regular analytical function of the wave number $k$ for all $k$
including the exceptional points. 
\end{abstract}

\pacs{02.40.Xx, 03.65.Nk, 03.65.Vf}

\maketitle

\section{Introduction}
Non-Hermitian Hamiltonians are widely used to describe open quantum
sistems in many fields of science
\cite{Moiseyev,joauadi,Kokoo,okolowicz,rotter,longhi,lefebvre}. For
continuum, $PT-$ symmetric, non Hermitian Hamiltonians on an infinite
line Bender et al.\cite{bender0,Bender} showed that the eigenvalue
spectrum is purely real when the strength of the non-Hermiticity is
small. Recently, Yogesh N et al. have investigated the signature of
$PT-$ symmetry breaking in coupled waveguides\cite{yogesh}. On the
other hand, the readings in a measuring device are real numbers which,
according to Quantum Mechanics, correspond to the points in the
spectrum of an operator that represents an observable.  This physical
condition is stated in mathematical form by demanding that the
spectrum of an operator representing an observable should be real
\cite{vNeumann1}.  When the observable is represented by a
self-adjoint operator the condition is automatically satisfied. But
the reality of the spectrum of an operator does not necessarily mean
that the operator is self-adjoint \cite{Mostafazadeh1, Fulop}.  This
non-equivalence of selfadjointness and the reality of the spectrum of
an operator was made evident by the discovery and subsequent
discussion of a large class of non-Hermitian $PT-$ symmetric
Schr\"odinger Hamiltonian operators with a complex valued potential
term but with real energy eigenvalues \cite{Bender,
  Mostafazadeh2,Cannata1}.  Even in the case of a radial Schr\"odinger
Hamiltonian with a real potential term, the reality of the energy
spectrum does not necessarily mean that the Hamiltonian is
self-adjoint. In many cases, the spectrum of a real non-self-adjoint
Hamiltonian is real but differs from the spectrum of self-adjoint ones
in two essential features. These are, the possible presence of
exceptional points and the possible presence of spectral
singularities.

Exceptional points already appear in the finite dimensional case of
non-Hermitian Hamiltonian matrices depending on a set of control
parameters
\cite{Heiss1,seven,Cavalli,FelixOlga,Thilagam,Dietz,moiseyev,kalita},
whereas spectral singularities are characteristic features of
Hamiltonians having a continuous energy spectrum
\cite{Kurasov1,Mostafazadeh3,Mostafazadeh4,Gusein,Longhi1,Andrianov1,
  samsonov,kalita,lefebvre}.  Hence, they are not possible for finite
dimensional operators.  Exceptional points in the real continuous
spectrum of a Schr\"odinger Hamiltonian with a real potential have
received much less attention than in the finite dimensional case
\cite{Andrianov2,Sokolov2,longhi}. With the purpose of clarifying the
topological nature of the exceptional points in the real, continuos
energy spectrum of a quantum system, in the following we will present
and discuss an example of exceptional points in a Hamiltonian with a
real potential of von Neumann-Wigner type. This potential is generated
from the eigenfunctions of a free particle Hamiltonian by means of a
four times iterated Darboux transformation when the transformation
functions are degenerated in the continuum of eigenfunctions of the
free particle Hamiltonian \cite{Matveev,vNeumann2}.

This paper is organized as follows: In section 2, we generated a
Hamiltonian $H[4]$ by means of a four times iterated and completely
degenerated Darboux transformation which has two exceptional points in
its real and continuous degenerated spectrum. In section 3, we compute
the Jost solutions of $H[4]$ normalized to unit probability flux at
infinity. In section 4, we show that at $k = \pm q$, the Wronskian of
the two unnormalized Jost solutions of $H[4]$ vanishes, this property
identifies these points as exceptional points in the spectrum of the
Hamiltonian $H[4]$. Section 5, is devoted to show that at the
exceptional points, the Hamiltonian $H[4]$ has a Jordan block matrix
representation and a Jordan cycle of generalized eigenfunctions. In
section 6, we show that the regular scattering solution vanishes at
exceptional points, while the irregular scattering solution has a
double pole at $k = \pm q$ and the scattering matrix $S(k)$ is a
regular function of $k$. In section 7, we show that the presence of an
exceptional point in the spectrum of $H[4]$ does not alter the
unitarity evolution of the regular time dependent wave function. The
pseudounitary evolution of the irregular time dependent wave function
is examined in section 8. A summary of the main results and
conclusions are presented in section 9.

\section{The potential $V[4]$}
A Hamiltonian that has two exceptional points in its real and
continuous spectrum may be generated by means of a four times iterated
and completely degenerated Darboux transformation
\cite{Nicolas1,Nicolas2}. Thus, and according with Crum's generalization of the
Darboux theorem\cite{Crum}, the function
\begin{equation}\label{cero1}
\psi[4]= \frac{W_{2}(\phi,\partial_{q}\phi,\partial^{2}_{q}\phi,
  \partial^{3}_{q}\phi,
  e^{\pm ikr})}{W_{1}(\phi,\partial_{q}\phi,\partial^{2}_{q}\phi,
\partial^{3}_{q}\phi)},
\end{equation}
is an eigenfunction of the radial Schr\"odinger equation with the potential
\begin{equation}\label{cero2}
V[4] = V_{0} - 2\frac{d^{2}}{dr^{2}}\ln W_{1}(\phi,\partial_{q}\phi, 
\partial^{2}_{q}\phi, \partial^{3}_{q}\phi).
\end{equation}

In these expressions, $W_{1}(\phi, \partial_{q}
\phi,... \partial^{3}_{q}\phi)$ is the Wronskian of the transformation
function $\phi(q,r)$ and its first derivatives with respect to $q$,
and $W_{2}(\phi, \partial_{q} \phi,... \partial^{3}_{q}\phi, e^{\pm
  ikr})$ is the Wronskian of the transformation function, $\phi(q,r)$,
its first three derivatives with respect to $q$ and $e^{\pm ikr}$
which is an eigenfunction of the free particle radial Hamiltonian
$H_{0}$ with eigenvalue $E = k^{2}$. The transformation function is
\begin{equation}\label{cero3}
\phi(q,r) = \sin (qr + \delta(q)),
\end{equation}
and $\partial_{q}\phi$ is shorthand for $\partial \phi/\partial q$. 
The phase shift $\delta(q)$ in the right side of eq.(\ref{cero3})
is a smooth, odd function of the wave number $q$.

 The Wronskian $W_{1}(\phi,\partial_{q}\phi,\partial^{2}_{q}
\phi,\partial^{3}_{q}\phi) = W_{1}(q,r)$ is readily computed from 
(\ref{cero3}) and the
 potential $V[4]$ is obtained from $W_{1}(q,r)$ and its derivatives
 with respect to $r$,
\begin{equation}\label{cero4}
V[4] =-2\frac{1}{W^{2}_{1}(q,r)} \Bigl(W^{''}_{1}(q,r)W_{1}(q,r) - W_{1}^{'2}(q,r)
\Bigr).
\end{equation}
An explicit expression for $W_{1}(q,r)$ is the following
\begin{eqnarray}\label{cero5}
W_{1}(q,r) &=&   
16(q\gamma)^{4} -
12(q\gamma)^{2} + 8(q^{3}\gamma_{2})(q\gamma) - 12 (q^{2}\gamma_{1})^{2}\cr 
&+& 24[(q^{2}\gamma_{1})(q\gamma) +(q\gamma)^{2}]\cos 2\theta  
+ 3\sin^{2} 2\theta \cr
&+& [16(q\gamma)^{3} - 12q\gamma - 12q^{2}\gamma_{1} -
4q^{3}\gamma_{2}]\sin 2\theta,
\end{eqnarray}
where
\begin{eqnarray}\label{cero6}
\theta(r) &=& qr + \delta(q), \hspace{0.3cm} \gamma(r) =
\partial_{q}\theta = r + \gamma_{0}, \cr 
\gamma_{0} &=& \partial_{q}\delta(q),\hspace{0.9cm} \gamma_{1} \ =
\partial^{2}_{q}\delta(q), \cr
\gamma_{2} &=& \partial^{3}_{q} \delta(q).
\end{eqnarray}

For large values of $r$, the dominant term in the right hand side of
eq.(\ref{cero5}) is $(q\gamma)^{4}$ which is positive and grows with
$r$ as $r^{4}$. Hence, for large values of $r$, $W_{1}(q,r)$ is a
positive and increasing function of $r$. However, if the phase
$\delta(q)$ is left unconstrained, $W_{1}(q,r)$ could take a negative
value at the origin of the radial coordinate $(r = 0)$, in which case
it should vanish for some positive, non-vanishing value of $r$, giving
rise to a singularity of the potential $V[4]$ at that point.

A necessary condition for the validity of the method of the Darboux
transformation is that the potential generated should not have any
singularities that are not present in the initial potential. In the
case under consideration, this condition means that the Wronskian
$W_{1}(q,r)$ should not vanish for any positive value of
$r$. Therefore, to avoid the appearence of singularities in $V[4]$ at
finite values of $r$, we will put the condition 
\begin{equation}\label{cero7}
W_{1}(q,0) > 0.
\end{equation}

In explicit form, we have
\begin{eqnarray}\label{cero8}
W_{1}(q,0) &=& 16\Bigl(q\frac{d\delta(q)}{dq}\Bigr)^{4} -
12\Bigl(q\frac{d\delta(q)}{dq} \Bigr)^{2} +
8\Bigl(q^{3}\frac{d^{3}\delta(q)}{d
  q^{3}}\Bigr)\Bigl(q\frac{d\delta(q)}{dq} \Bigr) \cr
&-&
12\Bigl(q^{2}\frac{d^{2}\delta(q)}{dq^{2}}\Bigr)^{2} +
24\Bigl[\Bigl(q^{2}\frac{d^{2}\delta(q)}{dq^{2}}\Bigr)\Bigl(q\frac{d\delta(q)}
  {dq}\Bigr) + \Bigl(q\frac{d\delta(q)}{dq}\Bigr)^{2}\Bigr]\cr
&\times& \cos
2\delta(q) +  3\sin^{2}2\delta(q) \cr &+& 
\Bigl[16\Bigl(q\frac{d\delta(q)}{dq}\Bigr)^{3} -
  12\Bigl(q\frac{d\delta(q)}{dq}\Bigr) -
  12\Bigl(q^{2}\frac{d^{2}\delta(q)}{dq^{2}}\Bigr) -
  4\Bigl(q^{3}\frac{d^{3}\delta(q)}{dq^{3}}\Bigr)\Bigr] \cr 
&\times&
\sin 2\delta(q).
\end{eqnarray}
Despite its formidable appearence, this equation may readily be integrated. 
First, to simplify the notation, we define a new function $t(q)$ as
\begin{equation}\label{cero9}
t(q):= \tan\delta(q),
\end{equation}
then
\begin{equation}\label{cero13}
\sin 2\delta(q) = \frac{2t(q)}{1 + t^2(q)} 
\end{equation}
and 
\begin{equation}\label{ceroo13}
\cos 2\delta(q) = \frac{1 - t^{2}(q)}{1 + t^2(q)}.
\end{equation}

Written in terms of $t(q)$, the condition of absence of
singularities in $V[4]$ takes the form
\begin{eqnarray}\label{cero10}
\frac{1}{4} \Bigl(1+t^{2}(q)\Bigr)^{2}Ẉ_{1}(q,0) &=&
\Bigl(-t(q) + qt_{q}(q)\Bigr)\Bigl[
3(-t(q) + qt_{q}(q)) + 6q^{2}t_{qq}(q) \cr
&+& 2q^{3}t_{qqq}(q)\Bigr] - 
 3q^{4}\Bigl(\frac{d}{dq}(-t(q) + qt_{q}(q))\Bigr)^{2},  
\end{eqnarray}
in this expression, $t_{q}$ is shorthand for $dt/dq$.

Now it is evident from (\ref{cero10}) that if $t(q)$ satisfies
\begin{equation}\label{cero11}
-t(q) + qt_{q}(q) = \beta, 
\end{equation}
the equation (\ref{cero10}) becomes an identity and the condition
(\ref{cero7}) is satisfied provided that

\begin{equation}\label{cero15}
W_{1}(q,0) = \frac{12\beta^{2}}{(1 + t^2(q))^{2}}.
\end{equation}

Integrating (\ref{cero11}) we get
\begin{equation}\label{cero12-0}
t(q) = \alpha q - \beta,
\end{equation}
according to equation (\ref{cero9})
\begin{equation}\label{cero12}
\delta(q) =\arctan(\alpha q - \beta),
\end{equation}
in these expressions $\alpha$ and $\beta$ are free parameters.

Once $\delta(q)$ is known as an explicit function of $q$, the
functions $\gamma_{0}, \ \gamma_{1}$ and $\gamma_{2}$ are obtained
from its first, second and third derivatives 

\begin{eqnarray}\label{cero14}
\gamma_{0} &=& \frac{d\delta(q)}{dq} = \frac{\alpha}{1 + (\alpha q -
  \beta)^{2}}, \cr 
\gamma_{1} &=& \frac{d^{2}\delta(q)}{dq^{2}} = -
\frac{2\alpha^{2}(\alpha q - \beta)}{(1 + (\alpha q - \beta)^{2})^{2}}, \cr 
\gamma_{2} &=& \frac{d^{3}\delta(q)}{dq^{3}} = - \frac{2\alpha^{3}(1
  - 3(\alpha q - \beta)^{2})}{(1 + (\alpha q - \beta)^{2})^{3}}.
\end{eqnarray}

With the help of these expressions and
(\ref{cero5}) we get for $W_{1}(q,r)$
\begin{eqnarray}\label{014}
W_1(q,r) &=& \frac{12\beta^2}{(1+(\alpha q - \beta)^2)^2} +
\frac{24\beta \alpha q}{(1+(\alpha q - \beta)^2)^2}(\cos 2qr -1 )\cr
 &+&  \frac{12\alpha q \left((\alpha q)^2 + \beta ^2 -
  1 \right)}{(1+(\alpha q - \beta)^2)^2} \sin 2qr  \cr 
&+& 16\Bigl[(qr)^4 
+ \frac{4\alpha q}{(1+(\alpha q - \beta)^2)}(qr)^3 
+  \frac{6(\alpha q)^2}{(1+(\alpha q - \beta)^2)^2}(qr)^2  \cr
&+& \frac{3(\alpha q)^3}{(1+(\alpha q - \beta)^2)^2}(qr)\Bigr] - 
12 \left[(qr)^2 + \frac{2\alpha q}{(1+(\alpha q -
    \beta)^2)}(qr)\right] \cr &+& 24 \left[ (qr)^2 
+  \frac{2\alpha q (1-\beta(\alpha q - \beta))}{(1+(\alpha q -
    \beta)^2)^2}(qr)\right]\cos 2(qr + \delta(q)) \cr 
& + & \Bigl[16\left((qr)^3 + \frac{3\alpha q}{(1+(\alpha q -
    \beta)^2)}(qr)^2 + \frac{3(\alpha q)^2}{(1+(\alpha q -
    \beta)^2)^2}(qr)\right)  \cr
&-&  12(qr)\Bigr]\sin 2(qr + \delta(q))  
+  3 \Bigl[\frac{1-6(\alpha q-\beta)^2 + (\alpha q -
    \beta)^4}{(1+(\alpha q - \beta)^2)^2}  \cr
&\times&  \sin^2 2qr + \frac{4(\alpha
    q-\beta)(1-(\alpha q - \beta)^2)}{(1+(\alpha q - \beta)^2)^2}\sin
  2qr \cos 2qr \Bigr].
\end{eqnarray}

From this expression, we verify that the Wronskian $W_{1}(q,r)$, for
$r = 0$, behaves as
\begin{equation}\label{cero15-0}
W_{1}(q,0) = \frac{12\beta^{2}}{(1 + (\alpha q - \beta)^{2})^2}.
\end{equation}
\begin{figure}[h]
\begin{center}
\includegraphics[width=300pt,height=170pt]{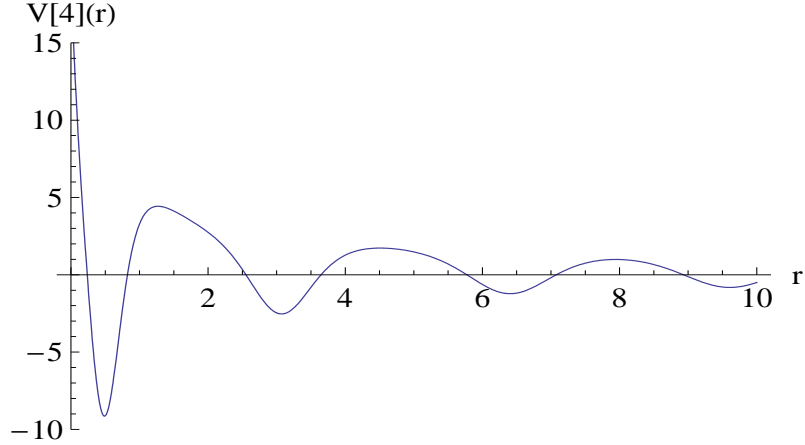}
\caption{This graph shows the potential $V[4](r)$ for the values of
  the parameters $\alpha = 1, \ \beta = 3$ fixed at $q = 1$}
\end{center}
\end{figure}

The asymptotic behaviour of $W_{1}(q,r)$ for large values of $r$ is
determined by the term $ (q\gamma)^{4} $ which grows as $r^{4}$ and is
the dominant term in the right hand side of eq.(\ref{cero5}),
with this result and equation (\ref{cero4}), we get
\begin{equation}\label{cero17}
\lim_{r \rightarrow \infty} V[4](r) \approx 8q\frac{\sin 2(qr +\delta(q))}{r} 
+ O\Bigl(r^{-2}\Bigr)
\end{equation}
Then, $V[4]$ is a potential of von Neumann-Wigner
type \cite{vNeumann2}. The figure 1 shows the behaviour of $V[4]$ as
function of $r$.

\section{The Jost eigenfunctions of $H[4]$}

The two linearly independent unnormalized Jost eigenfunctions of
$H[4]$, that belong to the energy eigenvalues $E = k^{2}$ and behave
as outgoing and incoming waves for large values of $r$ are given en
eq.(1). Notice that all terms in the last column of the Wronskian
$W_{2}(\phi, \partial_{q}\phi,...\partial^{3}_{q}\phi, e^{\pm ikr})$
are proportional to $e^{\pm ikr}$.

Hence, the Jost solutions of eq.(\ref{cero1}) take the form
\begin{equation}\label{cero50}
f^{\pm}(k,r) = \frac{1}{W_{1}(q,r)}w^{\pm}(k,r)e^{\pm ikr},
\end{equation}
where the function $w^{\pm}(k,r)$ is the reduced Wronskian defined as
\begin{equation}\label{cero51}
w^{\pm}(k,r)e^{\pm ikr} =
W_{2}(\phi,\partial_{q}\phi,...,\partial^{3}_{q}\phi, e^{\pm ikr}).
\end{equation}

From this definition, it follows that $w^{\pm}(k,r)$ is a complex
function of its arguments
\begin{equation}\label{cero52}
w^{\pm}(k,r) = u(k,r) \pm iv(k,r).
\end{equation}
A straightforward computation of the Wronskian $ W_{2}(\phi,
\partial_{q}\phi,...,\partial^{3}_{q}\phi, e^{\pm ikr})$ allowed us to
find the following explicit expressions for the functions
$u(k,r)$ and $v(k,r)$
\begin{eqnarray}\label{cero53}
u(k,r) & = & 12(k^4 + 6q^2 k^2 + q^4) \frac{\beta (\beta - 2\alpha
  q)}{(1 + (\alpha q - \beta)^2)^2} + 24 (k^4 - 4q^2 k^2 - q^4) \cr
&\times & \frac{\beta \alpha q}{(1+(\alpha q - \beta)^2)^2}\cos 2qr  +  
16 (k^2 - q^2)^2 \Bigl[(qr)^4  \cr
&+&  \frac{4 \alpha q}{(1+(\alpha q - \beta)^2)}(qr)^3  
+  \frac{6 (\alpha q)^2}{(1+(\alpha q - \beta)^2)^2}(qr)^2  \cr
&+&   \frac{3 (\alpha q)^3}{(1+(\alpha q - \beta)^2)^2}(qr)\Bigr] 
+  24\Bigl[(k^4 - 4q^2 k^2 - q^4)  \cr
&\times&  \left((qr)^2 + \frac{2 \alpha q}{(1+(\alpha q - \beta)^2)}(qr)
  \right) -  2(k^4 - q^4) \cr
&\times&   \frac{ (\alpha  q)^2(\alpha q - \beta)}
{(1+(\alpha q - \beta)^2)^2}(qr)\Bigr] 
\cos 2(qr + \delta (q))  +  \Bigl[ 16 (k^4 - q^4)  \cr
&\times&  \left((qr)^3 +
  \frac{3 \alpha q}{(1+(\alpha q - \beta)^2)}(qr)^2  
 +    \frac{3 (\alpha q)^2}
{(1+(\alpha q - \beta)^2)^2}(qr) \right) \cr
&-&  12(k^4 - 4q^2 k^2 - q^4)(qr)\Bigr]\sin 2(qr + \delta(q)) \cr
 & + & \Bigl[ 24(k^4 - q^4)\frac{(\alpha q)^3}{(1+(\alpha q -
    \beta)^2)^2} - 12(k^4 - 4q^2 k^2 - q^4) \cr
&\times&  \frac{\alpha q(1+(\alpha
    q)^2 - \beta ^2)}{(1+(\alpha q - \beta)^2)^2}\Bigr]\sin 2qr \cr 
& + & 3(k^4 + 6q^2 k^2 + q^4) \Bigl[ \frac{1-6(\alpha q-\beta)^2 +
    (\alpha q - \beta)^4}{(1+(\alpha q - \beta)^2)^2}\sin ^2 2qr
  \cr
& + & \frac{2(\alpha q-\beta)(1-(\alpha q -
    \beta)^2)}{(1+(\alpha q - \beta)^2)^2}\Bigr]\sin 2qr \cos 2qr
\end{eqnarray}
and
\begin{eqnarray}\label{cero54}
v(k,r) & = & 24qk \frac{\beta}{(1+(\alpha q - \beta)^2)^2}
\Bigl[(\beta ^2 -4\beta \alpha q -1)(k^2 + q^2)  \cr
&+&  (\alpha q)^2(q^2 + 5k^2) \Bigr]  +  \Bigl[24qk(k^2 + q^2) \cr
&\times&  \frac{\beta(\beta ^2 -
    4\beta\alpha q + (\alpha q)^2 +1)}{(1+(\alpha q - \beta)^2)^2} +
  96qk^3 \frac{\beta(\alpha q)^2}{(1+(\alpha q - \beta)^2)^2}
  \Bigr]\cr & \times &\cos 2qr + 64qk(k^2 - q^2) \Bigl[(qr)^3 +
  \frac{3\alpha q}{1+(\alpha q - \beta)^2}(qr)^2  \cr
&+&  \frac{3(\alpha
    q)^2}{(1+(\alpha q- \beta)^2)^2}(qr) \Bigr] - 24qk(k^2 +
q^2)(qr) + \Bigl[32qk(k^2 - q^2)  \cr 
&\times&   \left((qr)^3 + \frac{3\alpha
    q}{1+(\alpha q- \beta)^2}(qr)^2 + \frac{3(\alpha q)^2}{(1+(\alpha
    q- \beta)^2)^2}(qr) \right)  \cr
& + &  24qk(k^2 + q^2)(qr) \Bigr] \cos 2(qr + \delta (q))  \cr 
&+& \Bigl[96 q^3 k \left((qr)^2 + \frac{2\alpha q}{1+(\alpha q- \beta)^2}(qr) 
\right) \cr
 &+&  96qk(k^2 - q^2)\frac{(\alpha q)^2(\alpha q - \beta)}{(1+(\alpha q-
    \beta)^2)^2}(qr)\Bigr]\sin 2(qr+\delta (q)) \cr & + &
\Bigl[12qk(k^2+q^2)\frac{(\alpha q-\beta)^4+4\beta\alpha q
    -1}{(1+(\alpha q - \beta)^2)^2}  \cr
&-&  48qk(k^2-q^2)\frac{(\alpha
    q)^2}{(1+(\alpha q - \beta)^2)^2}\Bigr]\sin 2qr \cr &+& 12qk (k^2
+ q^2) \Bigl[\frac{1-6(\alpha q-\beta)^2 +(\alpha q-\beta)^4
  }{(1+(\alpha q-\beta)^2)^2} \sin 2qr \cos 2qr  \cr
&-& \frac{4(\alpha q-\beta)(1-(\alpha q-\beta)^2) }{(1+(\alpha
    q-\beta)^2)^2} \sin ^2 2qr \Bigr].
\end{eqnarray}

For large values of $r$, the asymptotic behaviour of $w^{\pm}(k,r)$ is
dominated by the highest power of $r$. From eqs.(\ref{cero53})
and (\ref{cero54}) we get

\begin{eqnarray}\label{cero56}
w^{\pm}(k,r) \approx 16(k^{2} - q^{2})^{2}(qr)^{4}
 \Bigl\{1 + O(r^{-1})\Bigr\}e^{\pm ikr}
\end{eqnarray}
and from eq.(\ref{014}) we get
\begin{equation}\label{cero57}
W_{1}(q,r) \approx 16(qr)^{4}\bigl[1 + O(r^{-1})\bigr],
\end{equation}
hence, for large values of $r$
\begin{equation}\label{cero58}
f^{\pm}(k,r) \approx (k^{2} - q^{2})^{2}\bigl[1 + O(r^{-1})\bigr]e^{\pm ikr}.
\end{equation}

The factor $(k^{2} - q^{2})^{2}$ is the flux of probability current at
infinity of the unnormalized Jost solutions.

Therefore, the Jost solutions of $H[4]$ normalized to unit
probability flux at infinity are
\begin{eqnarray}\label{cero59}
F^{\pm}(k,r) = \frac{f^{\pm}(k,r)}{(k^{2} - q^{2})^{2}} =
\frac{1}{(k^{2} -
  q^{2})^{2}}\frac{w^{\pm}(k,r)}{W_{1}(q,r)}
e^{\pm ikr}, \hspace{1 cm} k^{2} \ne q^{2}.
\end{eqnarray}

\section{Exceptional points in the spectrum of $H[4]$}

Each pair of linearly independent Jost solutions belongs to a point
$E_{k} = k^{2}$, with $k^{2} \neq q^{2}$, in the spectrum of $H[4]$.
At the point $E_{q} = q^{2}$, the two unnormalized Jost solutions
coalesce to give rise to a Jordan chain of two generalized bound state
eigenfunctions of $H[4]$\cite{Nicolas1}.

The Wronskian of the unnormalized Jost solutions is readily computed
from (\ref{cero50}) and (\ref{cero52}) 
\begin{eqnarray}\label{cero60}
W(f^{+}(k,r), f^{-}(k,r)) = -2ik(k + q)^{4}(k - q)^{4},
\end{eqnarray}

At the points $k = \pm q$, the Wronskian of the two unnormalized Jost
solutions of $H[4]$ vanishes, then the two unnormalized Jost solutions
are no longer lineraly independent, coalesce in one bound state
eigenfunction embeded in the continuum. This property identifies the
points $k = \pm q$ as exceptional points in the spectrum of the
Hamiltonian $H[4]$.

\section{Jordan cycle of generalized  eigenfunctions and Jordan block 
representation of $H[4]$}

The normalized Jost solutions of $H[4]$, as functions of the wave number
$k$, may be written as the sum of a singular and a regular part \cite{Nicolas1}
\begin{eqnarray}\label{cero18}
F^{\pm}(k,r) &=& \frac{\psi_{B}(q,r)e^{\mp i{\delta}(q)}}{(k - q)^{2}} + 
\frac{\chi_{B}^{\pm}(q,r)e^{\mp i{\delta}(q)}}{(k - q)}  +
\frac{\psi_{B}(- q,r)e^{\pm i{\delta}(q)}}{(k + q)^{2}} \cr 
&+& 
\frac{\chi_{B}^{\pm}(- q,r)e^{\pm i {\delta}(q)}}{(k + q)}  + 
h^{\pm}(k,r),
\end{eqnarray}
explicit expressions for the functions $\psi_{B}(q,r)$ and
$\chi_{B}^{\pm}(q,r)$ as functions of $r$, are
\begin{equation}\label{cero19-0}
\psi_{B}(q,r) = \frac{24 q^{2}}{W_{1}(q, r)}\{-2q^{2}\gamma^{2}\cos\theta + 
\bigl[q\gamma + q^{2}\gamma_{1}\bigr]\sin\theta + \sin^{2}\theta\cos\theta\}
\end{equation}
and 
\begin{equation}\label{cero20-0}
\chi_{B}^{\pm}(q,r) = \chi_{B}(q, r) \mp i\gamma_{0}\psi_{B}(q, r),
\end{equation}
where
\begin{eqnarray}\label{cuarentaisiete}
\chi_{B} (q,r) &=& \frac{8q}{W_{1}(q,r)}\bigl[-2q^{3}\gamma^{3}\sin\theta - 
3q^{2}\gamma^{2}\cos\theta + 3q\gamma\sin^{3}\theta \cr 
&-& \gamma_{2}q^{3}\sin\theta 
+ 3 \gamma_{1}\gamma q^{3}\cos\theta + 3\sin^{2}\theta\cos\theta \bigr].  
\end{eqnarray}
The functions $\gamma_{0}, \gamma_{1}$ and $\gamma_{2}$ are given in equations 
(\ref{cero14})
\begin{figure}[h]
\begin{center}
\includegraphics[width=300pt,height=170pt]{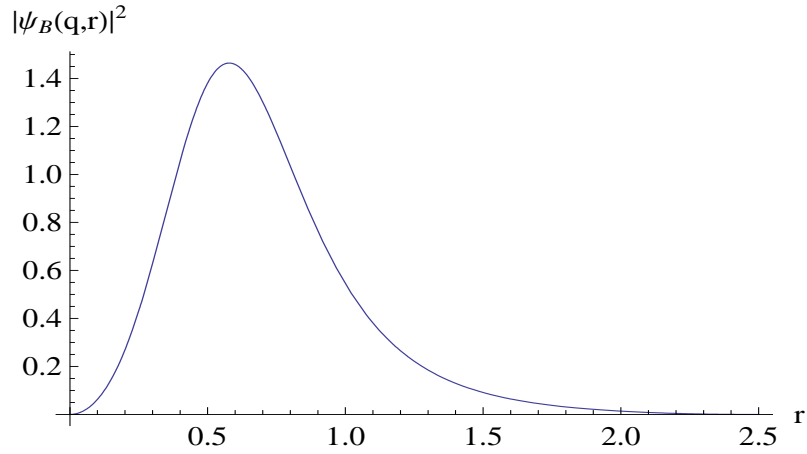}
\caption{This graph shows the bound state eigenfunction $\psi_{B}(q,
  r)$ as function of $r$ computed for $q = 1$ and the values of the
  parameters $\alpha = 1$ and $\beta = 3$}
\end{center}
\end{figure}

\begin{figure}[h]
\begin{center}
\includegraphics[width=300pt,height=170pt]{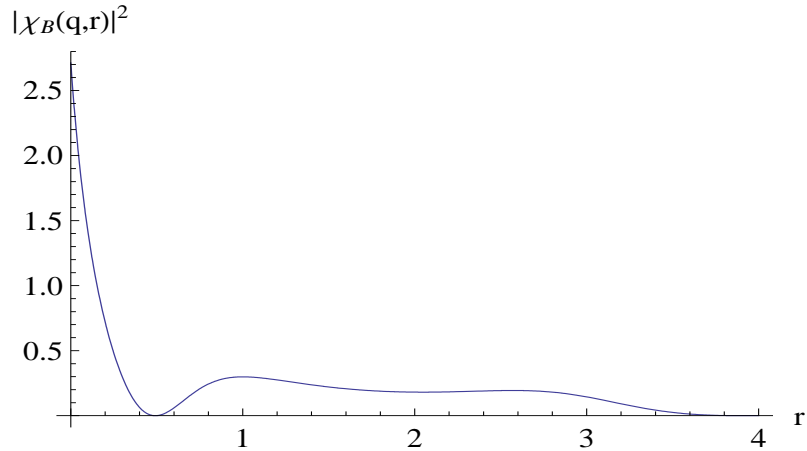}
\caption{The generalized eigenfunction $\chi_{B}(q,r)$, as a function
  of $r$ computed for $q=1$ and the values of the parameters $\alpha
  = 1$ and $\beta = 3$.}
\end{center}
\end{figure}
In the figures 2 and 3 we show the graphical representation of the
generalized eigenfunctions $\psi_{B}(q,r)$ and $\chi_{B}(q,r)$ as
function or $r$.

The functions $\psi_{B}(q,r)$ and $\chi_{B}(q,r)$ are the elements
of a Jordan cycle of generalized eigenfunctions of the Hamiltonian
that belong to the same point $k = q$ in its spectrum, and satisfy the
set of coupled equations
\begin{eqnarray}\label{cero19}
H[4]\psi_{B}(q,r) = q^{2}\psi_{B}(q,r)
\end{eqnarray}
and
\begin{eqnarray}\label{cero20}
H[4]\chi_{B}(q,r) = q^{2}\chi_{B}(q,r) + 2q\psi_{B}(q,r).
\end{eqnarray}
This Jordan cycle of generalized eigenfunctions is associated with a
Jordan block representation of the Hamiltonian $H[4]$. This
property is made evident if the equations (\ref{cero19}) and
(\ref{cero20}) are written in matrix form
\begin{eqnarray}\label{cero21}
\Bigl[H[4]{\bf 1_{2\times 2}}\Bigr]\Psi_{B}(q,r) = {\cal{H}}_{B}(q)\Psi_{B}(q,r),
\end{eqnarray}
where
\begin{eqnarray}\label{cero22}
\Psi_{B}(q,r) =\pmatrix{\psi_{B}(q,r) \cr
\chi_{B}(q,r) }
\end{eqnarray}
and 
\begin{eqnarray}\label{cero23}
{\cal{H}}_{B}(q) = \pmatrix{q^{2} & 0 \cr
2q & q^{2}
}.
\end{eqnarray}
From (\ref{cero21}), it is evident that the matrix
${\cal{H}}_{B}(q)$ is a matrix representation of the Hamiltonian
$H[4]$ in the two dimensional functional space spanned by the generalized
eigenfunctions $\{ \psi_{B}(q,r), \chi_{B}(q,r)\}$. 

This space is a subset of the rigged Hilbert space of continuous,
complex functions of the variables $(q,r)$ with continuous first and
second derivatives with respect to $r$, with $r$ in the semi-infinite
straight line $0\leq r < \infty$. Therefore, the two dimensional
subspace of functions spanned by the generalized eigenfunctions $\{
\psi_{B}(q,r), \chi_{B}(q,r)\}$ is in the domain of $H[4].$

The matrix ${\cal{H}}_{B}(q)$ is a Jordan block of $2\times
2$ \cite{Hernan} and can not be brought to diagonal form by means of a
similarity transformation with a unitary matrix\cite{Lancaster}.

The real non-symmetric matrix ${\cal{H}}_{B}(q)$ is $\eta -$
pseudo-Hermitian\cite{Mostafazadeh2}
\begin{equation}\label{cero24}
{\cal H}_{B}^{\dagger}(q) = \eta^{-1}{\cal H}_{B}(q)\eta
\end{equation}
with
\begin{eqnarray}\label{cero25}
{\eta} = \pmatrix{0 & 1 \cr
1 & 0
}.
\end{eqnarray}

 Hence, also the Hamiltonian operator 
$H[4]$ when acting on the functional space spanned by the
generalized eigenfunctions $\{ \psi_{B}(q,r), \chi_{B}(q,r)\}$ is also
$\eta-$pseudo-Hermitian.

\section{Scattering solutions  and the scattering matrix}
In this section it will be shown that the regular scattering solution
$\psi_{s}(k,r)$ vanishes at $k = \pm q$, and the irregular scattering
solution $\psi_{is}(k,r)$ has a double pole at the exceptional points $
k = \pm q$.

The regular scattering solution of $H[4]$ is given by
\begin{equation}\label{34-000}
\psi_{s}(k,r) = \frac{i}{2}\Bigl[F^{-}(k,r) - S(k)F^{+}(k,r)\Bigr], 
\hspace{0.5cm} k \neq q,
\end{equation}
where the scattering matrix $S(k)$ is
\begin{equation}\label{treintaicinco}
S(k) = \frac{f^{-}(k,0)}{f^{+}(k,0)} = \frac{w^{-}(k,0)}{w^{+}(k,0)} =
\frac{u(k,0) - iv(k,0)}{u(k,0) +i v(k,0)}.
\end{equation}
From the Jost solutions eq.(\ref{cero59}), we get
\begin{eqnarray}\label{54-n}
  \psi_{s}(k,r) &=& \frac{i}{2}\frac{1}{(k^2 -q^2)^2}\frac{1}{W_1(q,r)}
\frac{1}{w^{+}(k,0)}\Bigl[w^{+}(k,0)w^{-}(k,r)e^{-ikr} \cr
&-& 
w^{-}(k,0)w^{+}(k,r)e^{ikr}\Bigr].
\end{eqnarray}
a graphical representation of the regular scattering solution is shown
in figure 4.

\begin{figure}[h]
\begin{center}
\includegraphics[width=240pt,height=150pt]{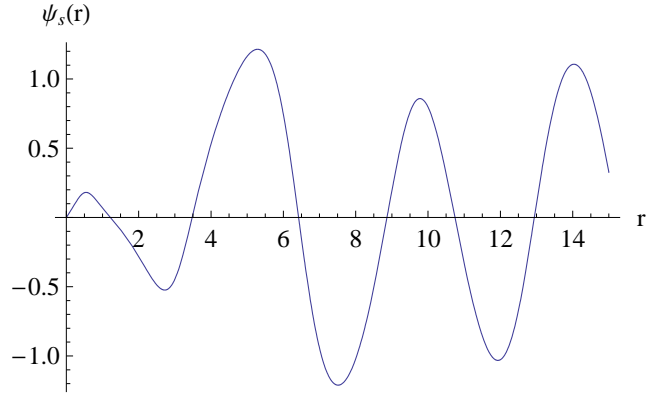}
\caption{The graph shows the regular scattering solution as function
  of $r$ for the values of the parameters in $V[4]$ and $k = 1.5$.}
\end{center}
\end{figure}

 In order to make explicit the properties of the scattering solutions
 as functions of $k$, for $k$ close to the exceptional points at 
$k = \pm q$, we will write the reduced Wronskian $w^{\pm}(k,r)$ as an
 expansion in powers of $(k - q)$

\begin{eqnarray}\label{56-n}
w^{\pm}(k,r) = u(k,r) \pm i v(k, r) = e^{\mp i\theta(q)}
\sum_{\ell = 0}^{4} w^{\pm}_{\ell} (q,r)(k - q)^{\ell},
\end{eqnarray}
explicit expressions for the coefficients $w^{\pm}_{\ell}(q,r)$, as
functions of their arguments are given in the appendix A.

Then, the terms in the right hand side of
eq.(\ref{54-n}) may also be expressed as an expansion in powers of 
$(k - q)$
\begin{eqnarray}\label{57-n}
 \psi_{s}(k,r) &=& \frac{i}{2}\frac{1}{(k^2 -q^2)^2}\frac{1}{W_1(q,r)}
\frac{1}{w^{+}(k,0)}\sum_{\ell = 0}^{4}\sum_{m =0}^{4}
\Bigl[w^{+}_{\ell}(q,0)w^{-}_{m}(q,r) \cr
&\times& e^{-i(k-q)r}
- w^{-}_{\ell}(q,0)w^{+}_{m}(q,r)e^{i(k-q)r}\Bigr](k - q)^{(\ell + m)},
\end{eqnarray}
In appendix B, we have shown that the first three terms in the 
summation in the right hand side of eq.(\ref{57-n}) vanish.

It follows that
\begin{eqnarray}\label{59-n}
\psi_{s}(k,r) &=&
\frac{i}{2}\frac{(k-q)}{(k+q)^{2}W_{1}(q,r)}\frac{1}{w^{+}(k,0)}\sum_{\ell
  = 3}^{8}\sum_{m=0}^{\ell}\Bigl[ w^{+}_{\ell -
    m}(q,0)w^{-}_{m}(q,r)e^{-i( k -q)r} \cr &-& w^{-}_{\ell - m}(q,0)w^{+}_{m}
  (q,r)e^{i(k -q)r}\Bigr] (k -q)^{\ell - 3} + O(k - q)
\end{eqnarray}
with the restriction $w_{\ell}^{\pm}(q,r) = 0,$ for $\ell\geq 5.$

Therefore, at the excepcional point, the regular scattering solution
vanishes,
\begin{equation}\label{60-n}
\psi_{s}(q, r) = 0.
\end{equation}
The energy eigenfunctions at the exceptional point are the generalized
bound state eigenfunctions $\psi_{B}(q,r)$ and $\chi_{B}(q,r)$.

Let us now turn our atention to the irregular scattering solution,
\begin{equation}\label{treintaicuatro}
\psi_{is}(k,r) = \frac{1}{2}\Bigl[F^{-}(k,r) + S(k)F^{+}(k,r)\Bigr], 
\hspace{0.5cm} k \neq q.
\end{equation}

\begin{figure}[h]
\begin{center}
\includegraphics[width=240pt,height=150pt]{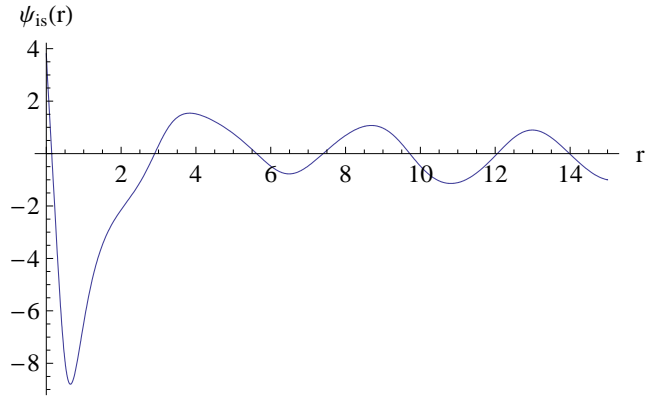}
\caption{The graph shows the irregular scattering solution as function
  of $r$ for the values of the parameters in $V[4]$ and $k = 1.5$.}
\end{center}
\end{figure}
In this case, we have
\begin{eqnarray}\label{54-nn}
  \psi_{is}(k,r) &=& \frac{1}{2}\frac{1}{(k^2 -q^2)^2}\frac{1}{W_1(q,r)}
\frac{1}{w^{+}(k,0)}\Bigl[w^{+}(k,0)w^{-}(k,r)e^{-ikr} \cr
&+& 
w^{-}(k,0)w^{+}(k,r)e^{ikr}\Bigr].
\end{eqnarray}
 Notice that, when the expression (\ref{56-n}) is substituted for
 $w^{\pm}(k,r)$ in (\ref{54-nn}) we obtain an expansion of the right
   hand side of (\ref{54-nn}) in powers of $(k -q)$ similar to
     (\ref{57-n}), but now the terms proportional to
     $w^{\pm}_{0}(q,0)w^{\pm}_{0}(q,r)$,  $w^{\pm}_{0}(q,
     0)w^{\pm}_{1}(q,r)$ and $w^{\pm}_{1}(q,0)w^{\pm}_{1}(q,r)$ are added.
     Hence, the singular terms in the expressions
     (\ref{cero18}) for the Jost solutions add instead of
     cancelling which makes $\psi_{is}(k,r)$ singular at the
     exceptional points. Hence, the irregular scattering solution
     $\psi_{is}(k,r)$, as a function of $k$, has a pole of second
     order at the exceptional points.

Substitution of the expressions (\ref{56-n}) for
$w^{\pm}(k,r)$ in (\ref{54-nn}) and multiplying the result times $(k -
q)^{2}$ and taking the limit $k \rightarrow q $ gives,
\begin{equation}\label{sesentaiseis}
\lim_{k \rightarrow q} \left(k - q\right)^{2}\psi_{is}(k,r) = \psi_{B}(q,r).
\end{equation}

Therefore, the bound state eigenfunction $\psi_{B}(q,r)$ is the
residue of second order of the irregular scattering solution
$\psi_{is}(k,r)$ at $k = q$.

From equations (\ref{cero53}) and (\ref{cero54}) for the funciones
  $u(k,r)$ and $v(k,r)$ evaluated in $ r = 0$ we get
\begin{eqnarray}\label{60-000}
u(k,0) &=& \frac{12\beta}{(1 + (\alpha q - \beta)^2)^2}
 \left[\beta k^4 + (6\beta - 20\alpha q) k^2 q^2 + (\beta - 4\alpha q)q^{4}
\right]
\end{eqnarray}
and
\begin{eqnarray}\label{61-000}
v(k,0) &=& \frac{48\beta  q k}{(1 + (\alpha q - \beta)^2)^2} \cr
 &\times& \left[(\beta^{2} - 4\beta\alpha q + 5\alpha^{2}q^{2})k^{2}
 + (\beta^{2} - 4\beta\alpha q + \alpha^2 q^2)q^{2}\right],
\end{eqnarray}
substitution of the equations (\ref{60-000}) and (\ref{61-000}) in
(\ref{treintaicinco}) allows to write the following expression for the
scattering matrix $S(k)$
\begin{equation}\label{62-00}
S(k) = \exp{2i\Delta(k)}
\end{equation}  
where the phase shift $\Delta(k)$ is 
\begin{equation}\label{62-000}
\Delta (k) = -\arctan \frac{4q k \left[(\beta^{2} - 4\beta\alpha q + 
5\alpha^{2}q^{2})k^{2} + (\beta^{2} - 4\beta\alpha q + \alpha^{2}q^{2})q^{2}
\right]}{\beta k^{4} + (6\beta -20\alpha q)q^{2}k^{2} + (\beta -4\alpha q)q^{4}},
\end{equation}
\begin{figure}[h]
\begin{center}
\includegraphics[width=240pt,height=150pt]{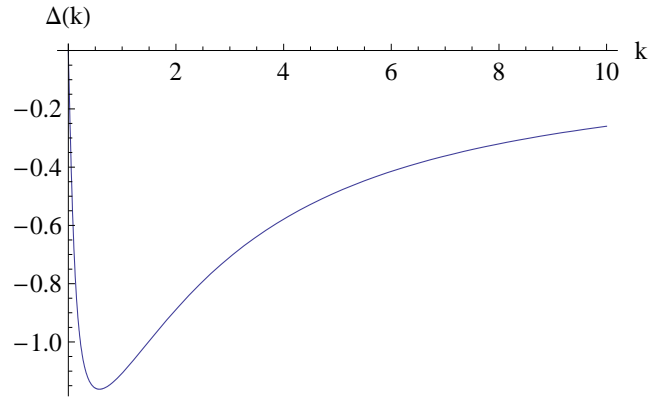}
\caption{The phase shift $\Delta(k)$ as function of the wave number
  $k$ for $q=1$ and the values of the parameters $\alpha =$ 1 and $\beta =$ 3.}
\end{center}
\end{figure}
the figure 6 shows the graphical representation of the phase shift.

The scattering matrix is a regular analytical function of the wave
number $k$, including the exceptional points, it does
not have any singularity or pole.

\section{Unitarity time evolution  of the regular scattering eigenfunctions 
of $H[4]$}

The regular time dependent wave functions built as a linear
combination of the regular scattering eigenfunctions of $H[4]$ is
\begin{equation}\label{cero63}
\Psi_{r}(r,t) = \int C(k)e^{-i\frac{k^{2}(t - t_{0})}{\hbar}}
\psi_{s}(k,r) dk.
\end{equation}
This is a solution of the time dependent Schr\"odinger equation
\begin{equation}\label{cero64}
i\hbar \frac{\partial \Psi_{r}(r,t)}{\partial t} = H[4]\Psi_{r}(r,t),
\end{equation}
as it may be verified by substitution of (\ref{cero63}) in (\ref{cero64}).

The coefficient $C(k)$ occurring in the right hand side of eq.(\ref{cero63}) 
is a quadratically integrable function of $k$.

Since the regular scattering eigenfunctions vanishes at the
exceptional point
\begin{equation}\label{cero65}
\psi_{s}(q,r) = 0,
\end{equation}
it does not contribute to the integral in the right hand side of
eq.(\ref{cero63}).

Therefore, the presence of an exceptional point in the spectrum of
$H[4]$ does not alter the unitarity evolution of the regular time
dependent wave function.

\section{Pseudounitarity time evolution of the irregular scattering 
eigenfunctions}

The two generalized eigenfunctions $\psi_{B}(q,r)$ and $\chi_{B}(q,r)$
belong to the same spectral point, $E_{q} = q^{2}$, in consequence,
they evolve in time together. Hence, it should be convenient to
introduce a matrix notation to deal with the two together,
\begin{equation}\label{cero69}
\Psi (r,t) = C(q,t)\Psi_{B}(q,r),
\end{equation}
where $\Psi_{B}(q,r)$ is the two component vector of the doublet
\begin{eqnarray}\label{cero70}
\Psi_{B} (q,r)  = \pmatrix{\psi_{B}(q,r) \cr
\chi_{B}(q,r) }
\end{eqnarray}
and $C(q,t)$ is the $2\times 2$ matrix of time dependent coefficent of
the wave functions $\psi_{B}(q,r)$ and $\chi_{B}(q,r)$. 

Substitution of $\Psi(r,t)$ in the time dependent
Schr\"odinger equation gives the following set of coupled equations
written in matrix form
\begin{equation}\label{cero71}
i\hbar \frac{\partial C(q,t)}{\partial t}\Psi_{B}(q,r) = C(q,t) H[4]
{\bf 1}_{2\times 2}\Psi_{B}(q,r) = C(q,t){\cal H}_{B}(q)\Psi_{B}(q,r),
\end{equation}
making abstraction of $\Psi_{B}(q,r)$, we obtain
\begin{equation}\label{cero72}
i\hbar\frac{\partial C(q,t)}{\partial t} = C(q,t){\cal H}_{B}(q),
\end{equation}
where
\begin{eqnarray}\label{cero73}
{\cal H}_{B}(q) = \pmatrix{q^{2} & 0 \cr
2q & q^{2} }.
\end{eqnarray}

The matrix ${\cal H}_{B}(q)$ is a representation of the Hamiltonian
$H[4]$ in the two dimensional functional space spanned by the
generalized eigenfunctions $\{\psi_{B}(q,r), \chi_{B}(q,r)\}$.

Integrating eq.(\ref{cero72}) we get
\begin{equation}\label{cero74}
C(q,t) = e^{-i\frac{1}{\hbar}{\cal H}(q)t},
\end{equation}
writing ${\cal H}_{B}(q)$ in explicit form in (\ref{cero74}), and
computing the exponential, we obtain
\begin{eqnarray}\label{cero75}
C(q,t) = e^{-i\frac{q^{2}t}{\hbar}}\pmatrix{1 & 0 \cr
-i\frac{2qt}{\hbar} & 1 }.
\end{eqnarray}

Substitution of the expressions (\ref{cero75}) for $C(q,t)$ and
(\ref{cero70}) in (\ref{cero69}) gives
\begin{equation}\label{cero76}
\psi_{B}(r,t) = e^{-i\frac{q^{2}t}{\hbar}}\psi_{B}(q,r)
\end{equation}
and
\begin{equation}\label{cero77}
\chi_{B}(r,t) = e^{-i\frac{q^{2}t}{\hbar}}\chi_{B}(q,r) - 
i\frac{2q}{\hbar}t e^{-\frac{iq^{2}t}{\hbar}}\psi_{B}(q,r).
\end{equation}

\section{Summary and conclusions}
An example of exceptional points in the continuous spectrum of a real,
pseudo-Hermitian Hamiltonian $H[4]$ of von Neumann-Wigner type is
presented and discussed.  The Hamiltonian $H[4]$ and the free particle
Hamiltonian $H_{0}$ are isospectral. In the general case, to each
point in this continuous spectrum, correspond two linearly independent
Jost solutions which behave at infinity as incoming and outgoing
waves. Clearly in both cases, this continuous spectrum is doubly
degenerate. However, here we have shown that in the continuous
spectrum of $H[4]$ there are two exceptional points at wave numbers $k
= \pm q$, these exceptional points are associated with a double pole
in the normalization factor of the Jost eigenfunctions normalized to
unit flux at infinity. At the exceptional points, the two unnormalized
Jost eigenfunctions are no longer linearly independent and coalesce to
give rise to Jordan cycles of lenght two of generalized quadratically
integrable bound states eigenfunctions embedded in the continuum and a
Jordan block representation of the Hamiltonian $H[4]$.  The regular
scattering eigenfunction vanishes at the exceptional point and the
irregular scattering eigenfunction has a double pole with a second
order residue equal to the bound state eigenfunction in the continuum
and a Jordan bock representation of the Hamiltonian. In consequence,
the time evolution of the regular scattering eigenfunction is unitary,
while the time evolution of the irregular scattering eigenfunction is
pseudounitary. The scattering matrix $S(k)$ is a regular function of
$k$ at the exceptional points, that is, the Jordan cycle of
generalized bound states eigenfunctions of $H[4]$ in the continuum is
not associated with a pole of the scattering matrix.

\subsection{Acknowledgments}
This work was partially supported by CONACyT M\'exico under Contract
No. 132059 and by DGAPA-UNAM Contract No. PAPIIT:IN113712.

\begin{appendix}
\section{}
The dependence of $w^{\pm}(k,r)$ on $k$ is readily made evident by expanding
the reduced Wronskian defined in eq.(\ref{56-n}) in powers of $(k-q)$
\begin{equation}\label{E-A1}
  w^{\pm}(k,r) = {\rm e}^{\mp i\theta}\sum_{\ell =0}^{4}w^{\pm}_{\ell}(q,r) 
(k - q)^{\ell},  
\end{equation}
the coefficients of the powers of $(k-q)$ in this expansion are
\begin{eqnarray}\label{E-A2}
w _{0}^{\pm}\left( q,r\right) = 4q^{2}W_{1}(q,r)\psi_{B}(q,r)
\end{eqnarray}
\begin{eqnarray}\label{E-A3}
w_{1}^{\pm }\left( q,r\right) = 4qW_{1}(q,r)\Bigl[\psi_{B}(q,r) +
  q\chi_{B}(q,r) \mp iq\gamma\psi_{B}(q,r)\Bigr]
\end{eqnarray}
\begin{eqnarray}\label{E-A4}
w_{2}^{\pm}\left( q,r\right)  &=& -2W_{1}(q,r)\psi_{B}(q,r)\cos^{2}\theta + 
6q W_{1}(q,r) \chi_{B}(q,r) \cr
&+& 16q^{2}\Bigl\{4q^{4}\gamma^{4} - 3q^{2}\gamma^{2} - 3q^{4}\gamma_{1}^{2} + 
2q^{4}\gamma\gamma_{2} + 3\sin^{2}\theta\cos^{2}\theta\Bigr\}\cos\theta \cr
&\mp& i2W_{1}(q,r)\Bigl\{(2q\gamma + q^{2}\gamma_{1})\psi_{B}(q,r) + 
2q^{2}\gamma\chi_{B}(q,r)\Bigr\}
\end{eqnarray}

\begin{eqnarray}\label{E-A5}
\omega^{\pm}_{3}(q,r) &=& 2W_{1}(q,r)\chi _{B}(q,r) - \frac{4}{q} 
W_{1}(q,r)\psi _{B}(q,r) \cr
&+& 8q\left( (8q^4\gamma^4 - 18q^2\gamma^2 + 3q^3 \gamma_1 \gamma -6q^4
\gamma_1^2)\cos\theta \right. \cr
&+& \left. (8q^3\gamma^3 + 3q\gamma -12q^2 \gamma_1 + 2q^3\gamma_2^2)
\sin^3\theta \right. \cr
&-& \left. 12\bigl(2q^2\gamma^2 + q^3\gamma_1\gamma-1\bigr)\right)
\sin^2\theta\cos\theta  \cr
&\mp & i \Bigl[ 4q\gamma W_{1}(q,r)\chi_{B}(q,r) + 2\bigl(\gamma + 
\gamma_1\bigr)  W_{1}(q,r)\psi _{B}(q,r)   \cr
&-&  8q\left( (12q^3\gamma^3 - 6q^2\gamma_1 -3q^3 \gamma_2 )
\sin^2\theta\cos\theta \right.  \cr
&+& \left.  12(q^2\gamma^2 + 3q^3\gamma_1\gamma )\cos^2\theta\sin
\theta - 12q^3\gamma^3\cos\theta \right)\Bigr] 
\end{eqnarray}

\begin{eqnarray}\label{E-A6}
\omega ^{\pm}_{4}(q,r) &=& \Bigl[ -\frac{1}{q^2} W_{1}(q,r)\psi _{B}(q,r)
\cos\theta + 8(4q^3\gamma^3 + q^3\gamma_2)\sin\theta\cos\theta  \cr
& + &  16q^4\gamma^4 - 36 q^2\gamma^2 + 12q^2\gamma_1(2q\gamma -
q^2\gamma_1) +36\sin^2\theta\cos^2\theta  \cr 
 &-&  48q^3\gamma_1\gamma\sin^2\theta \Bigr]\cos\theta \pm i  
\Bigl[ -\frac{1}{q^2} W_{1}(q,r)\psi _{B}(q,r)\cos\theta  \cr
 &+& 8(4q^3\gamma^3 + q^3\gamma_2)\sin\theta\cos\theta + 
16q^4\gamma^4 - 36 q^2\gamma^2  \cr
 &+&  12q^2\gamma_1(2q\gamma -q^2\gamma_1) + 36\sin^2\theta\cos^2\theta 
- 48q^3\gamma_1\gamma\sin^2\theta \Bigr] \cr
&\times& \sin\theta 
\end{eqnarray}

An expansion of the reduced Wronskian, $w^{\pm}(k,r)$, about the point
$k = - q$ is readily obtained from eq. (\ref{E-A1}). From time
reversal invariance of the Hamiltonian plus boundary conditions  we get
\begin{equation}\label{e-7}
f^{+}(-k,r) = f^{-}(k,r)
\end{equation}
and
\begin{equation}\label{E-A7}
w^{\pm}(k,r) = w^{\mp}(-k,r).
\end{equation}

Hence, from (\ref{E-A1})
\begin{equation}\label{E-A8}
w^{\pm}(k,r) = e^{\mp i\theta(q)}\sum_{\ell = 0}^{4} (- 1)^{\ell} w^{\pm}_{\ell}(q,r) 
(k + q)^{\ell},
\end{equation}
the coefficents $w^{\pm}_{\ell} (q,r)$ are explicitly given in
eqs.(\ref{E-A2} - \ref{E-A6}) as entire functions of
$q$. Their domain of analiticity is trivially extended to negative
values of the argument $q$ just by changing $q$ by $-q$ in
(\ref{E-A2} - \ref{E-A6}). Then it may be verified that
\begin{equation}\label{E-A9}
(-1)^{\ell}w^{\pm}_{\ell}(q,r) = w^{\pm}_{\ell}(-q,r).
\end{equation}

Therefore,
\begin{equation}\label{E-A10}
w^{\pm}(k,r) = e^{\pm i\theta(-q)}\sum_{\ell = 0}^{4}(-1)^{\ell}w^{\pm}_{\ell}(|q|,r) 
(k + q)^{\ell}, \hspace{0.4cm} k < 0.
\end{equation}

\section{}
In this appendix it will be shown that the regular scattering
eigenfunction vanishes at the exceptional points $k = \pm q.$

The regular scattering solution, eq.(\ref{cero59}), is
\begin{eqnarray}\label{B1}
  \psi_{s}(k,r) &=& \frac{i}{2}\frac{1}{(k^2 -q^2)^2}\frac{1}{W_1(q,r)}
\frac{1}{w^{+}(k,0)}\Bigl[w^{+}(k,0)w^{-}(k,r)e^{-ikr} \cr
&-& w^{-}(k,0)w^{+}(k,r)e^{ikr}\Bigr].
\end{eqnarray}

With the reduced Wronskian $w^{\pm}(k,r)$ as an expansion in powers of
$(k-q)$, eq.(\ref{E-A1}), we get
\begin{eqnarray}\label{B2}
\psi_{s}(k,r) &=&
\frac{i}{2}\frac{1}{(k^2 -q^2)^2}\frac{1}{W_1(q,r)}
\frac{1}{w^{+}(k,0)}\sum_{\ell = 0}^{8}\sum_{m=0}^{\ell}\Bigl[ w^{+}_{\ell -
    m}(q,0)w^{-}_{m}(q,r) \cr
&\times&  e^{-i( k -q)r} - w^{-}_{\ell - m}(q,0)w^{+}_{m}
  (q,r)e^{i(k -q)r}\Bigr] (k -q)^{\ell},
\end{eqnarray}
with the restriction $w_{\ell}^{\pm}(q,r)=0$ for $\ell \geq 5.$

Writing the first three terms in the summation, corresponding to
$\ell=0,1$ and $2$, and using the expansion of $e^{-i(k-q)r}$, we get
\begin{eqnarray}\label{B3}
\psi_{s}(k,r) &=&
\frac{i}{2}\frac{1}{(k^2 -q^2)^2}\frac{1}{W_1(q,r)}
\frac{1}{w^{+}(k,0)}\{\zeta_{0}(q,r)  \cr
&+&  \zeta_{1}(q,r)(k-q)+\zeta_{2}(q,r)(k-q)^2 +O((k-q)^{3})  \cr
&+&  \sum_{\ell= 3}^{8}\sum_{m=0}^{\ell}\Bigl[ w^{+}_{\ell -m}(q,0)w^{-}_{m}(q,r)
e^{-i( k -q)r}  \cr
&-&  w^{-}_{\ell - m}(q,0)w^{+}_{m}(q,r)e^{i(k -q)r}\Bigr] (k -q)^{\ell} \},  
\end{eqnarray}
where the functions $\zeta_{0}(q,r), \zeta_{1}(q,r)$ y
$\zeta_{2}(q,r)$ are given by
\begin{equation}\label{B4}
\zeta_{0}(q,r)= w_{0}^{+}(q,0)w_{0}^{-}(q,r)-c.c.
\end{equation}
\begin{eqnarray}\label{B5}
\zeta_{1}(q,r) &=& -iw_{0}^{+}(q,0)w_{0}^{-}(q,r)r+w_{1}^{+}(q,0)w_{0}^{-}(q,r)\cr
&+& w_{0}^{+}(q,0)w_{1}^{-}(q,r) - c.c.
\end{eqnarray}
\begin{eqnarray}\label{B6}
\zeta_{2}(q,r) &=&-\frac{1}{2}w_{0}^{+}(q,0)w_{0}^{-}(q,r)r^2 - iw_{1}^{+}(q,0)
w_{0}^{-}(q,r)r \cr 
&-& iw_{0}^{+}(q,0)w_{1}^{-}(q,r)r + w_{2}^{+}(q,0)w_{0}^{-}(q,r)r\cr
&+& w_{1}^{+}(q,0)w_{1}^{-}(q,r) + w_{0}^{+}(q,0)w_{2}^{-}(q,r)-c.c.
\end{eqnarray}
in these expressions, c.c. is shorthand for complex conjugate.

From the eq.(\ref{E-A2}), $w_{0}^{+}(q,r)=w_{0}^{-}(q,r)=w_{0}(q,r)$, where 
$w_{0}(q,r)$ is the real function
\begin{equation}\label{B7}
w_{0}(q,r)=4q^2 W_{1}(q,r) \psi_{B}(q,r)
\end{equation} 
then, from eq.(\ref{B4}),  
\begin{equation}\label{B8}
\zeta_{0}(q,r)=0.
\end{equation}

From eq.(\ref{E-A3}) we note that the imaginary part of $w_{1}^{\pm}(q,r)$ is
\begin{equation}\label{B9}
w_{1}^{\pm}(q,r)= \mp 4 W_{1}(q,r)q^{2}\gamma \psi_{B}(q,r),
\end{equation}
hence, $\zeta_{1}(q,r)$ is written as
 \begin{eqnarray}\label{B10}
 \zeta_{1}(q,r)&=&-2iw_{0}(q,0)w_{0}(q,r)r-8iq^2\gamma_{0}W_{1}(q,0)\psi_{B}(q,0)w_{0}(q,r)\nonumber\\
 & &+8i q^2\gamma w_{0}(q,0)W_{1}(q,r)\psi_{B}(q,r)
 \end{eqnarray}
and using (\ref{B7}) we get
\begin{equation}\label{B11}
\zeta_{1}(q,r)=2iw_{0}(q,0)w_{0}(q,r)(-r-\gamma_{0}+\gamma).
\end{equation}
But, from eq.(\ref{cero6}), $\gamma=r+\gamma_{0}$, then
\begin{equation}\label{B13}
\zeta_{1}(q,r)=0.
\end{equation}

The function $\zeta_{2}(q,r)$ is computed from the expressions for $w_{1}^{\pm}(q,r)$ and $w_{2}^{\pm}(q,r)$, we get
\begin{eqnarray}\label{B12}
\zeta_{2}(q,r)&=&-8iqW_{1}(q,0)[\psi_{B}(q,0)+q\chi_{B}(q,0)]w_{0}(q,r)r\nonumber\\
&-&8iqW_{1}(q,r)[\psi_{B}(q,r)+q\chi_{B}(q,r)]w_{0}(q,0)r\nonumber\\
&+&32iq^2W_{1}(q,0)[\psi_{B}(q,0)+q\chi_{B}(q,0)]W_{1}(q,r)q\gamma \psi_{B}(q,r)\nonumber\\
&-&32iq^2W_{1}(q,r)[\psi_{B}(q,r)+q\chi_{B}(q,r)]W_{1}(q,0)q\gamma_{0} \psi_{B}(q,0)\nonumber\\
&-&4iW_{1}(q,0)[(2q\gamma_{0}+q^2\gamma_{1})\psi_{B}(q,0)+2q^2\gamma_{0}\chi_{B}(q,0)] w_{0}(q,r)\nonumber\\
&+&4iW_{1}(q,r)[(2q\gamma+q^2\gamma_{1})\psi_{B}(q,r)+2q^2\gamma\chi_{B}(q,r)] w_{0}(q,0).
\end{eqnarray} 

Using eq.(\ref{B7}) for $w_{0}(q,r)$, we get
\begin{eqnarray}\label{B14}
\zeta_{2}(q,r)&=&32 i q^3 W_{1}(q,r)W_{1}(q,0)\left\lbrace 2 \psi_{B}(q,r) \psi_{B}(q,0)\right.\nonumber\\
&+& \left. q \left[\psi_{B}(q,r) \chi_{B}(q,0)+\psi_{B}(q,0) \chi_{B}(q,r)\right]\right\rbrace (-r-\gamma_{0}+\gamma).
\end{eqnarray} 
But, $\gamma=r+\gamma_{0}$, then 
\begin{equation}\label{B15}
\zeta_{2}(q,r)=0.
\end{equation}
With these results, the regular scattering solution $\psi_{s}(k,r)$, as function of $k$, close to the exceptional point $k=q$, is 
\begin{eqnarray}\label{B16}
\psi_{s}(k,r) &=& \frac{i}{2}\frac{1}{(k+q)^2}\frac{k-q}{W_1(q,r)}
\frac{1}{w^{+}(k,0)} \sum_{\ell= 3}^{8}\sum_{m=0}^{\ell}\Bigl[
  w^{+}_{\ell -m}(q,0)w^{-}_{m}(q,r)\cr &\times& e^{-i( k -q)r} -
  w^{-}_{\ell - m}(q,0)w^{+}_{m}(q,r)e^{i(k -q)r}\Bigr] (k
-q)^{\ell-3} \cr &+& O((k-q)).
\end{eqnarray}

\end{appendix}

\end{document}